\documentclass[journal]{IEEEtran}

\ifCLASSINFOpdf
\else
   \usepackage[dvips]{graphicx}
\fi
\usepackage{url}

\usepackage{graphicx}
\usepackage{amsmath}
\usepackage[bb=boondox]{mathalfa}
\usepackage{caption}

\usepackage{tikz}
\usetikzlibrary{shapes,arrows}

\tikzstyle{block} = [draw,text width=8em, minimum height=3em, text centered, fill=black!5, rectangle]
\tikzstyle{pinstyle} = [pin edge={to-,thin,black}]

\begin{document}

\title{Learning Multi-instrument Classification with Partial Labels}

\author{Amir Kenarsari Anhari}

\markboth{}
{Shell \MakeLowercase{\textit{et al.}}: Bare Demo of IEEEtran.cls for IEEE Journals}
\maketitle

\begin{abstract}
Multi-instrument recognition is the task of predicting the presence or absence of different instruments within an audio clip. A considerable challenge in applying deep learning to multi-instrument recognition is the scarcity of labeled data. OpenMIC is a recent dataset containing 20K polyphonic audio clips. The dataset is weakly labeled, in that only the presence or absence of instruments is known for each clip, while the onset and offset times are unknown. The dataset is also partially labeled, in that only a subset of instruments are labeled for each clip. 

In this work, we investigate the use of attention-based recurrent neural networks to address the weakly-labeled problem. We also use different data augmentation methods to mitigate the partially-labeled problem. Our experiments show that our approach achieves state-of-the-art results on the OpenMIC multi-instrument recognition task.
\end{abstract}


\IEEEpeerreviewmaketitle

\section{Introduction}
\IEEEPARstart{M}{usic} information retrieval (MIR) is an interdisciplinary field which is related to various disciplines including signal processing, information retrieval, machine learning, multimedia engineering, musicology, and digital humanities \cite{muller2015fundamentals}. Recent advances in machine learning (ML) models and artificial intelligence (AI) have changed traditional approaches in MIR \cite{langkvist2014review}. 

Musical instrument recognition is a well-known problem in the MIR field. Humans can easily recognize the instruments used in a piece of music by combining multiple perception modalities, but it still is a challenging task for a machine to recognize the instrument. While several classical machine learning and deep learning methods have been developed for single instrument recordings with great success \cite{lostanlen_extended_2018, han_sparse_2016, herrera2003automatic}, recognizing instruments in polyphonic recordings is still a challenging problem. The task is more difficult as the signal is a superposition of different sources/instruments and they have different timbre characteristics. Therefore, most of the isolated instrument recognition techniques that have been proposed in the literature are inappropriate for polyphonic music signals.  

Instrument recognition has many use cases in the MIR field. For example, it is desirable to include the instrument information in the audio tags to allow users to search for a piece of music with the instrument that they want \cite{fu2010survey}. In addition, it can be used to improve the performance of other MIR tasks. For instance, it can improve the performance of tasks such as source separation and automatic music transcription by knowing the type and number of instruments in a recording.

Some previous works such as Kitahara et al. \cite{kitahara_instrument_2006} experimented with the extraction of spectral and temporal features followed by Latent Discriminant Analysis (LDA). More recent works have focused on applying deep neural networks. For instance, Li et al. \cite{li} applied a Convolutional Neural Network (CNN) architecture similar to popular CNNs used in computer vision to learn audio features from raw audio on the MeledyDB dataset. Hung et. al \cite{hung2018} proposed several methods for frame-level instrument recognition by using constant-Q transform (CQT) features on the MusicNet dataset. However, most of these datasets are relatively small, while modern machine learning methods such as deep learning often benefit greatly from larger datasets for training.

Audioset \cite{gemmeke2017audio} is an ontology and human-labeled dataset of sounds extracted from YouTube videos. Audioset is the first dataset that achieves a similar scale as ImageNet \cite{deng2009imagenet} dataset in computer vision. The current (v1) version consists of more than 2 million 10-second audio clips organized into a hierarchical ontology, annotated with 527 sound labels, leading to 5{,}800 hours of audio in total. Since the dataset is intended for general purpose audio classification, it is suitable for a variety of problems related to audio and music \cite{gao2018learning} and can be used as an embedding extractor for similar tasks.  

In 2018, an open dataset for multi-instrument called \emph{OpenMIC} \cite{openmic} was released. OpenMIC consists of 20K 10-second audio clips extracted from Free Music Archive (FMA). OpenMIC provides an opportunity for researchers to investigate a larger and more diverse set of sounds instead of being limited to a small dataset with a limited number of instruments. The data creators also released the embedding vector for each audio extracted from the model trained on Audioset. Table \ref{table:dataset} lists the open polyphonic datasets used for multi-instrument recognition with a more detailed description for each of one of them.

\begin{table}[ht]
\centering 
\begin{tabular}{c c c} 
\hline\hline 
Name & Size & Number of Instruments \\ [0.5ex] 
\hline 
MeledyDB \cite{medleydb} & 122 & 80 \\ 
MusicNet \cite{musicnet}& 330 & 11 \\
OpenMIC-2018 \cite{openmic} & 20000 & 20 \\ [1ex]
\hline 
\end{tabular}
\caption{Comparison of available collections for multi-instrument recognition.} 
\label{table:dataset} 
\end{table}

One challenge of using OpenMIC for multi-instrument recognition is that the dataset is a weakly-labeled dataset (WLD) \cite{kumar2016audio}. That is, for each audio clip only the presence or absence of an instrument is known, while the onset and offsets times are not specified. In contrast to WLD, strongly-labeled data (SLD) refers to the data labeled with both the instrument labels and onset and offset times. However, strongly-labeled datasets are usually limited to a relatively small size as the onset and offset labeling is a time-consuming process, Therefore, strongly-labeled datasets usually limit the performance of neural network models that require a relatively large dataset to train a good model. To address the weakly labeled dataset problem, Gururani et al. \cite{attention} proposed an attention-based model to attend to specific segments in the audio clip. They evaluated the model on OpenMIC dataset and found out that attention improves the performance of the model compared to random forests, fully connected neural networks, and recurrent neural networks.

In this paper, we train a multi-instrument classifier on the OpenMIC dataset. The contribution of this work include the following:

\begin{itemize}
    \item Recurrent neural network with an attention layer is introduced to extract temporal features from a dataset.
    \item Modified cross-entropy loss function is proposed for multi-instrument recognition that attends more towards harder instruments.
    \item The impact of different data augmentation mechanisms on multi-instrument recognition is studied.
\end{itemize}

This paper is organized as follows. Section \ref{sec:modelling} presents the proposed deep neural network architecture, loss function, and different augmentation techniques. Section \ref{sec:exp} shows the experimental setup and results. Conclusions are drawn in Section \ref{sec:conclusion}. 

\section{Proposed Method}
\label{sec:modelling}
We present here the network architecture, train loss, and the augmentation strategies used in this paper.
\bigskip

\textbf{Notation:} We denote the number of instruments and number of train samples by \emph{C} and \emph{N}, respectively. The train data is represented by \(D=\{(S^{(1)}, \mathbf{y^{(1)}}),...,(S^{(N)}, \mathbf{y^{(N)}})\}\), where $S^{(i)}$ is the $i^{th}$ audio clip and \(\mathbf{y^{(i)}}=[y_{1}^{(i)},...,y_{C}^{(i)}] \subseteq \{-1, 0, 1\}^{C}\). For a given audio clip $i$ and instrument class $c$, value of $y_{c}^{(i)}=1$ (-1 and 0 respectively) means that the instrument is present (absent and unknown respectively).  

\subsection{Model}
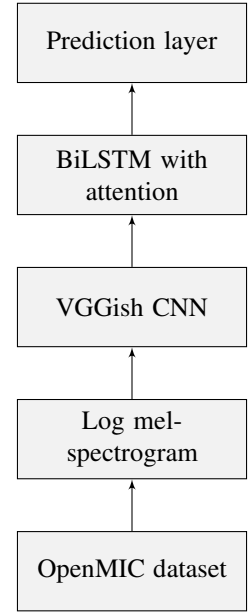
\begin{figure}[h!]
    \centering
    \begin{tikzpicture}[auto, node distance=5em,>=latex']
    \node [block, name=openmic] (openmic) {OpenMIC dataset};
    \node [block, above of=openmic] (mel) {Log mel-spectrogram};
    \node [block, above of=mel] (vgg) {VGGish CNN};
    \node [block, above of=vgg] (lstm) {BiLSTM with attention};
    \node [block, above of=lstm] (sigmoid) {Prediction layer};
    \draw [->] (openmic) -- (mel);
    \draw [->] (mel) -- (vgg);
    \draw [->] (vgg) --  (lstm);
    \draw [->] (lstm) -- (sigmoid);
\end{tikzpicture}
    \caption{Model architecture.}
    \label{fig:model}
\end{figure}

The proposed model is an attention-based Bi-directional Long Short-term Memory (BiLSTM) which takes the high-level features from the VGGish model proposed in \cite{vggish}. The model is pictured in Figure \ref{fig:model}. The VGGish CNN layer learns the local features from the log mel-spectrogram of the raw audio clip and the BiLSTM layer learns the temporal features. Finally, the attention layer aggregates the predictions in different timestamps and produces the final prediction. The BiLSTM layer has 64 units in each direction and we set both the dropout rate and recurrent dropout to 0.2 to reduce the risk of over-fitting \cite{gal2016theoretically}. These configurations are chosen empirically. The attention layer is modeled by following the formulation described in \cite{kong2018audio}.

\subsection{Objective Loss}
We use a modified version of cross-entropy, known as focal loss that is designed for object detection \cite{lin2017focal}. The focal loss applies a modulating term to the cross-entropy loss to focus on learning hard instruments and to prevent the easily classified examples to overwhelm the cross-entropy loss. We write the focal loss as:

\begin{equation}
L(p_t)= -{\alpha}_t(1-p_t)^\gamma \log(p_t)
\end{equation}

where:
\begin{equation}
p_t = \left\{ \,
\begin{IEEEeqnarraybox}[][c]{l?s}
\IEEEstrut
p & if $y=1$, \\
1-p & otherwise.
\IEEEstrut
\end{IEEEeqnarraybox}
\right.
\label{eq:example_left_right1}
\end{equation}

In the above, \(p \in [0, 1]\) is the model's output that shows the probability the instance belongs to an instrument. Note that in our formulation since we are dealing with unclassified labels, we only consider $y=1$ and $y=-1$ in our loss function and labels with $y=0$ are not included in the calculation of the loss value. The focusing parameter \(\gamma\) adjusts the rate at which easy examples are down-weighted and \(\alpha\) is a balancing multiplier for the loss function.

In our experiment, we use Adam optimizer \cite{kingma2014adam}, with initial learning rate at $5e^{-4}$. We set the focal loss parameters, $\alpha$ and $\gamma$ to 0.75 and 2, respectively.

\subsection{Data Augmentation}
OpenMIC is partially-labeled. That is, some instruments are not labeled and cannot be used for training. To account for this, we utilize a data augmentation method, known as mix-up to provide similar but different examples to our model \cite{zhang2017mixup}. Each new sample is constructed as a linear interpolation of two random examples from the training set with their labels and is given by:

\begin{equation}
\begin{aligned}
\bar{s} = \lambda{s^{(i)}} + (1-\lambda){s^{(j)}} \\
\bar{y} = \lambda\mathbf{y^{(i)}} + (1-\lambda)\mathbf{y^{(j)}}
\end{aligned}
\end{equation}

where $(s^{(i)}, \mathbf{y^{(i)}})$ and $(s^{(j)}, \mathbf{y^{(j)}})$ are train examples drawn randomly from the training set and weight \(\lambda \in [0, 1]\). We followed the recommendation in \cite{zhang2017mixup} to draw the weight from Beta distribution ($\lambda \sim Beta(\alpha, \alpha)$). The hyper-parameter $\alpha$ is set to 0.2 in our experiments.

We use another data augmentation technique to expose our model to different examples. Since the inputs to the BiLSTM can have different lengths, we can easily create new examples by concatenating them. Specifically, We achieve this by simply picking random examples during the training and concatenating them. Furthermore, the mixed label can be easily generated by performing an element-wise \emph{OR} operation between them.

\section{Experiments and Results}
\label{sec:exp}
We present here some experiments conducted to develop the system. OpenMIC provides the split for the training and test data. We use the model presented in section \ref{sec:modelling} and use the model proposed in \cite{attention} as a baseline model. The experiments are performed on a workstation with an NVIDIA GPU 1060 with 6GB memory. All algorithm is implemented in Python by using the machine learning framework TensorFlow \cite{tensorflow}. In the experiments, the number of the epoch was set to 200. The input batch size was set to 32.

\begin{figure}
    \centering
    \includegraphics[scale=0.5]{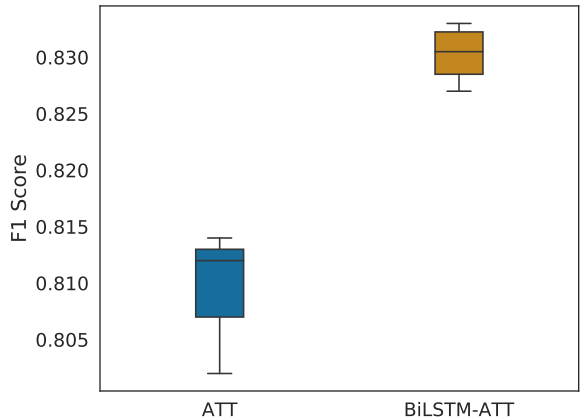}
    \caption{Macro-averaged F1 score for different architectures.}
    \label{fig:f1_score}
\end{figure}

\subsection{Dataset}
We experiment on the OpenMIC dataset \cite{openmic}. OpenMIC contains 20K 10-second audio clips extracted from FMA, and annotated with 20 instrument labels. It includes the bottleneck feature vectors extracted from the embedding layer representation of the VGGish CNN proposed in \cite{vggish}. The embedding layer produces one feature vector per second which is post-proceeded by a principal component analysis (PCA) to remove the correlations and only the first 128 PCA components are kept. The 20K audio clips were split into disjoint train and test sets. We used 15{\%} of train set as validation set.

\subsection{Metrics}
In these experiments, we measure the F1 score. Accuracy is not used as a metric due to the highly unbalanced nature of the multi-instrument recognition problem. Macro F1 is evaluated by averaging per-instrument F1, while micro F1 is evaluated on the results of all the recordings over all the instruments. For computing these metrics, we use a confidence threshold of 0.5, i.e., if the prediction of the model is greater than 0.5, the prediction is taken as positive.

True positive (TP), where both the ground truth and the system prediction are positive; false negative (FN) is where the ground truth is positive but the system prediction is negative; false positive (FP), where the ground truth is negative but the system prediction is positive; true negative (TN), where both ground truth and system prediction are negative. The precision (P) and recall (R) are as follows \cite{mesaros2016metrics}

\begin{equation}
    P = \frac{TP}{Tp+FP}
\end{equation}

\begin{equation}
    R = \frac{TP}{Tp+FN}
\end{equation}

In addition, the F1 score is defined as

\begin{equation}
    F1=\frac{2PR}{P+R}
\end{equation}

For these experiments, we report per-instrument metrics and also macro-metrics, where we macro-average the per-class metrics.

\subsection{Comparison with Other Methods}
\begin{figure}
    \centering
    \includegraphics[scale=0.52]{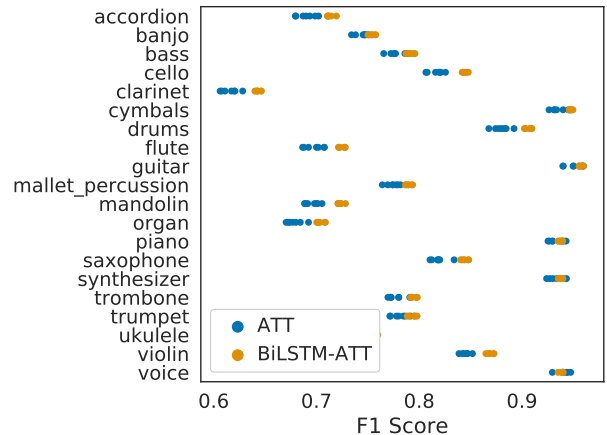}
    \caption{Instrument-level F1 score for different architectures.}
    \label{fig:instrument_f1_score}
\end{figure}

We compare our method with the most recent method \cite{attention}, which uses attention with dense layers. Figure \ref{fig:f1_score} shows the macro-average F1 score for our proposed method (BiLSTM-ATT) and the method proposed in \cite{attention} which we refer to as ATT. We also report the F1 score for each instrument in Figure \ref{fig:instrument_f1_score} for both methods. We see an overall performance gain in the F1 score comparing to the ATT model. One explanation for this performance gain is that the BiLSTM layer can model the temporal structure in the examples and the attention layer aggregates the predictions in each timestamp to provide a more accurate prediction for each instrument in the audio clip. 

\section{Conclusion}
\label{sec:conclusion}
This paper presents a deep framework to learn a multi-instrument classifier with partial labels. We propose an attention-based network architecture to address the WLD problem in OpenMIC dataset. Our experiments show that data augmentation such as mixup can improve the overall performance of the model. We show that using focal loss introduced originally for object detection helps the model to learn harder examples. In the future, we will investigate other approaches in data augmentation to improve the overall performance of the system.

\bibliographystyle{IEEEtran}
\bibliography{references}

\end{document}